\documentclass{mem}
\usepackage{natbib}\usepackage{txfonts}\usepackage{balance}
\usepackage{graphicx}
\usepackage[a4paper]{hyperref}
\idline{75}{1}
\begin{document}
\def\teff{$T\rm_{eff }$}
\def\kms{$\mathrm {km s}^{-1}$}
\def\gtorder{\mathrel{\raise.3ex\hbox{$>$}\mkern-14mu
    \lower0.6ex\hbox{$\sim$}}}
\def\ltorder{\mathrel{\raise.3ex\hbox{$<$}\mkern-14mu
    \lower0.6ex\hbox{$\sim$}}}

\title{
Galactic Bars in Cosmological Context 
}

   \subtitle{}

\author{
Isaac \,Shlosman\inst{1,2}
          }

  \offprints{I. Shlosman}

\institute{
JILA, University of Colorado --
Box 440,
Boulder, CO 80309-0440, USA 
\and
Department of Physics \& Astronomy,
University of Kentucky,
Lexington, KY 40506-0055, USA 
\email{shlosman@pa.uky.edu}
}

\authorrunning{Shlosman }

\titlerunning{Galactic Bars in Cosmological Context}

\abstract{
Galactic disks can form in asymmetric potentials of the assembling 
dark matter (DM) halos, giving
rise to the first generation of gas-rich bars. Properties of these bars
differ from canonical bars analyzed so far. Moreover, rapid disk growth 
is associated
with the influx of clumpy DM and baryons along the large-scale filaments.
Subsequent interactions between this substructure and the disk can trigger
generations of bars, which can explain their ubiquity in the Universe.
I provide a brief summary of such bar properties and argue that they fit 
naturally within the broad cosmological context of a hierarchical 
buildup of structure in the universe.
}
\maketitle{}

\section{Introduction}

Among various approaches to galaxy formation, two stand out --- those that
deal with isolated and open systems. In the former {\it modus operandi,}
angular momentum, $J$, and mass of a system are conserved. Thus, for example,
one deals with a finite reservoir of gas, galaxy interactions are treated
in maximally-controled experiments, and, in most such cases, the amount of
substructure present is quite limited, if it exists at all. The modeled 
galaxies quickly use up the available gas and show a single splash of gas 
influx to fuel the central starburst or AGN activity. As a result,
there are numerous difficulties in reproducing the observed phenomena in
isolated systems: they start with simplified initial conditions, e.g.,
an axial symmetry (in disk galaxies) which is broken spontaneously by the bar 
instability. In many ways these systems are too idealized to explain the 
evolutionary tracks of real galaxies, but they serve as excellent educational 
tools. The perennial question about these systems remains: 
why should nature create `equilibrium' systems and
immediately `destroy' them by various dynamical instabilities?

\begin{figure*}[]
\begin{center}
\resizebox{3.3cm}{!}{\includegraphics[clip=true]{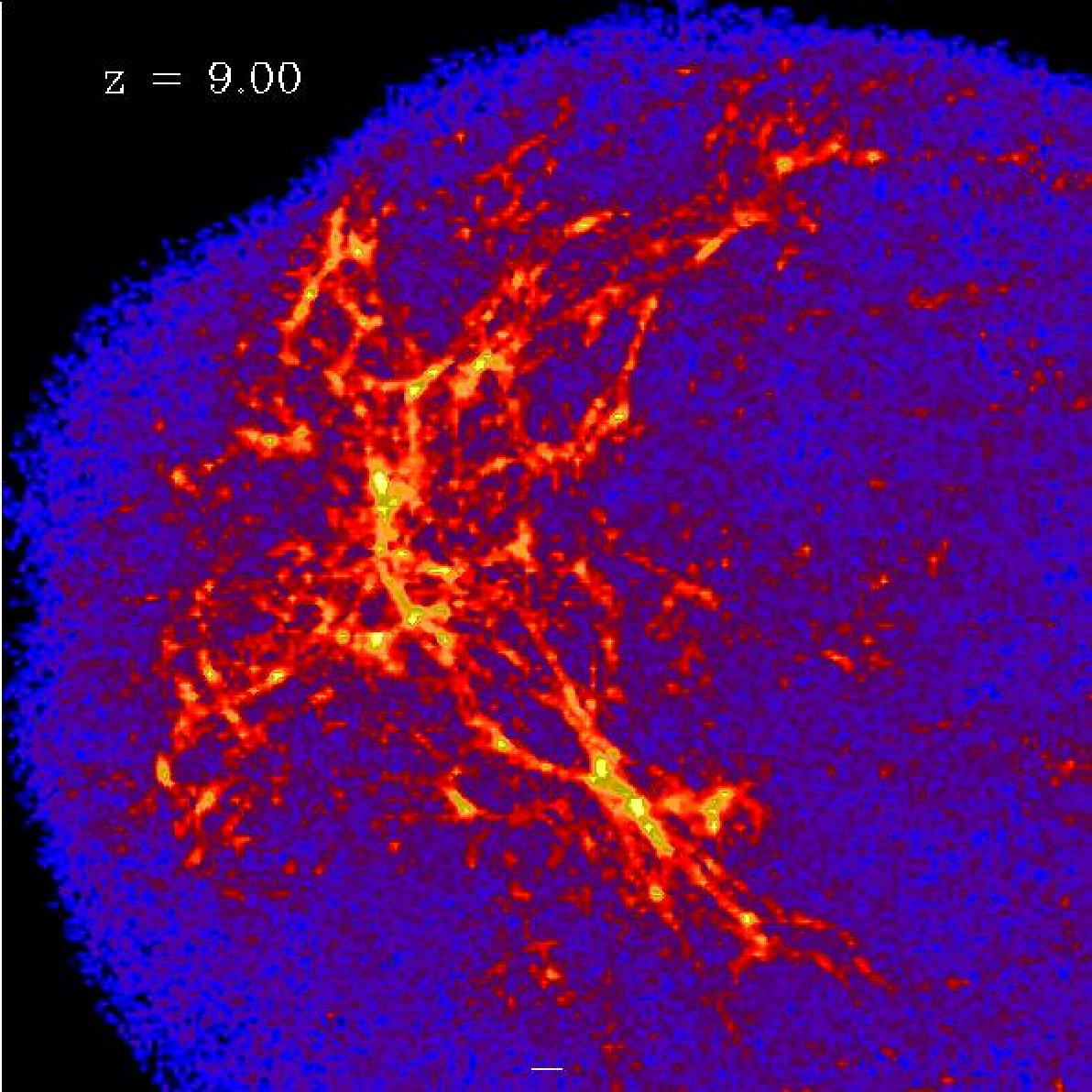}}
\resizebox{3.3cm}{!}{\includegraphics[clip=true]{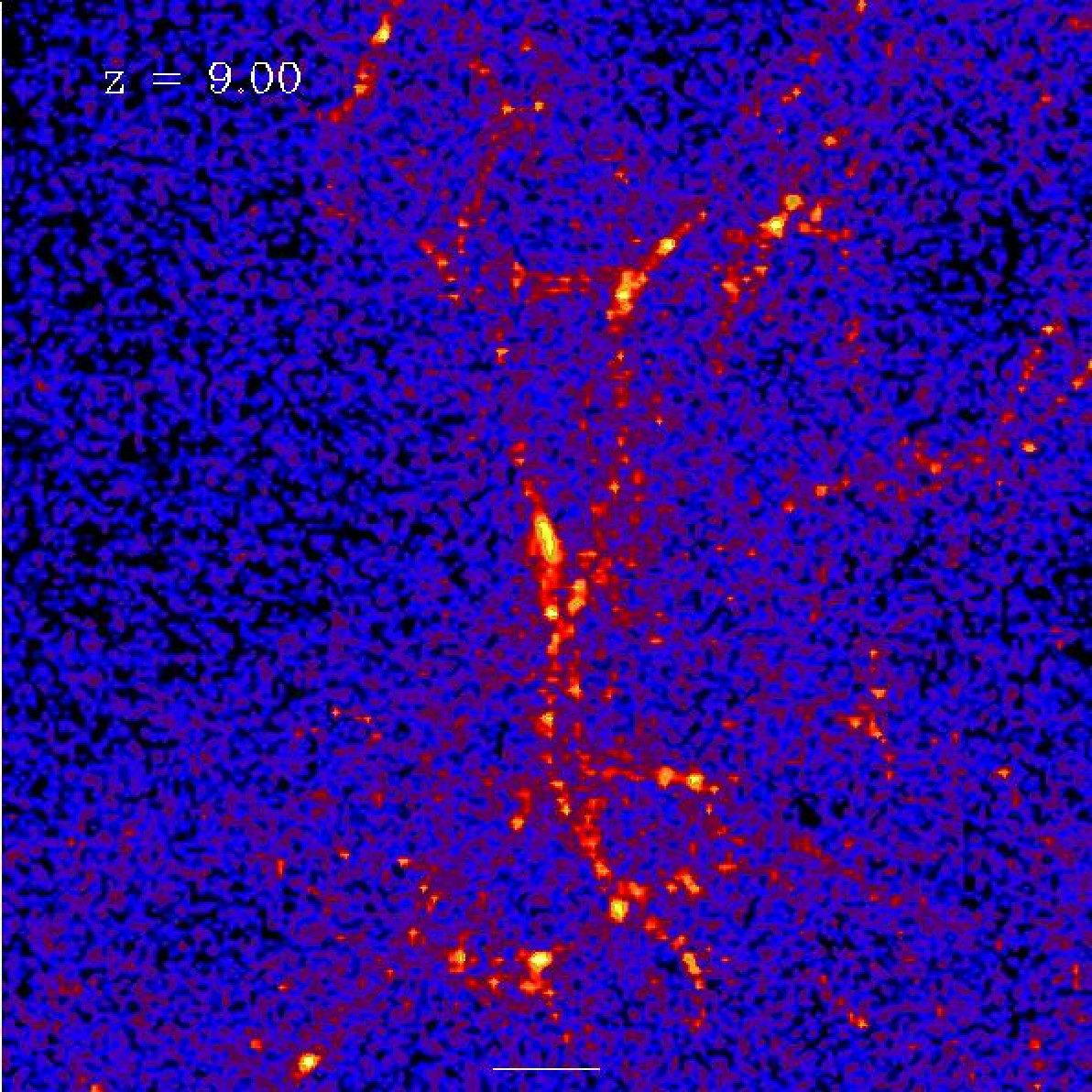}}
\resizebox{3.3cm}{!}{\includegraphics[clip=true]{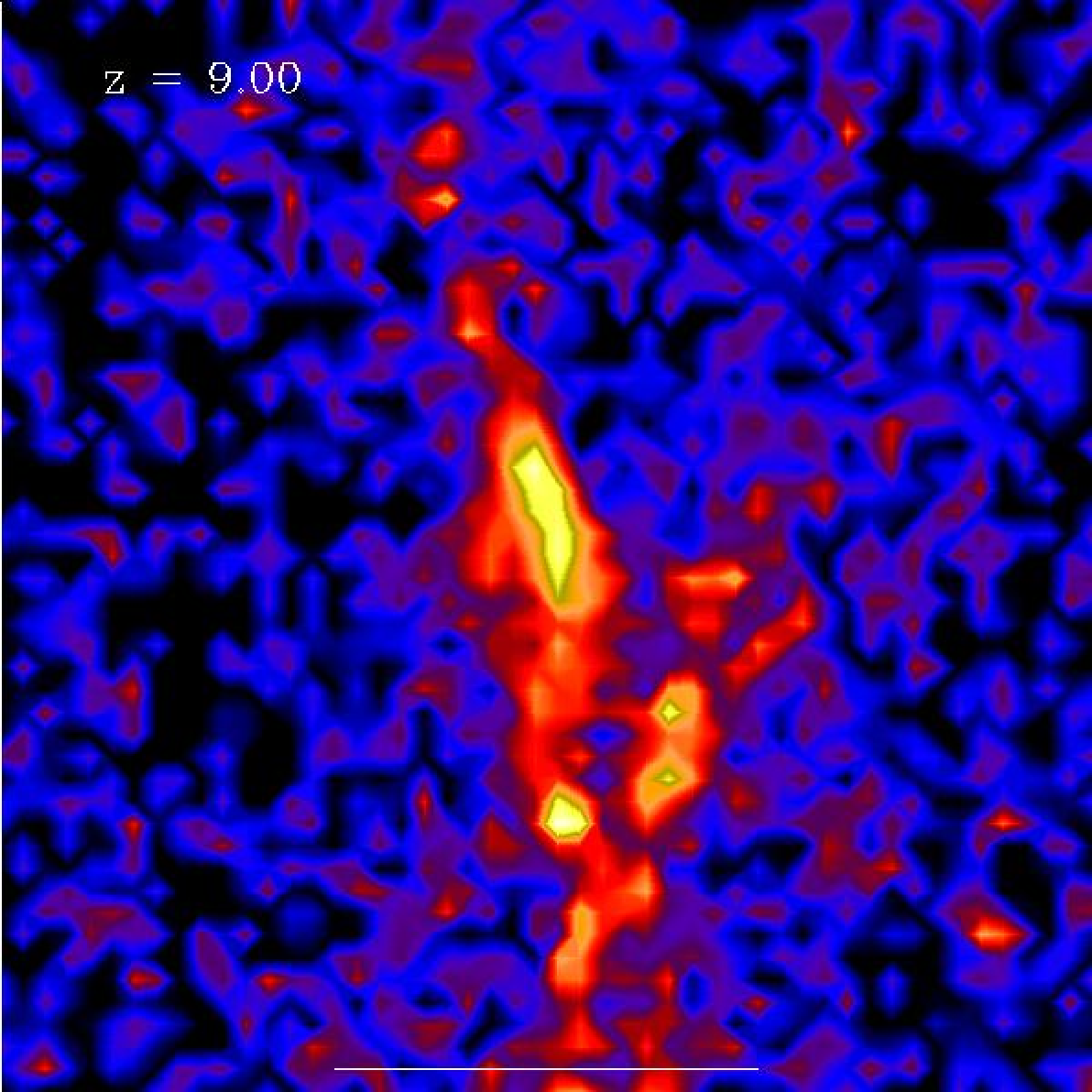}}
\resizebox{3.3cm}{!}{\includegraphics[clip=true]{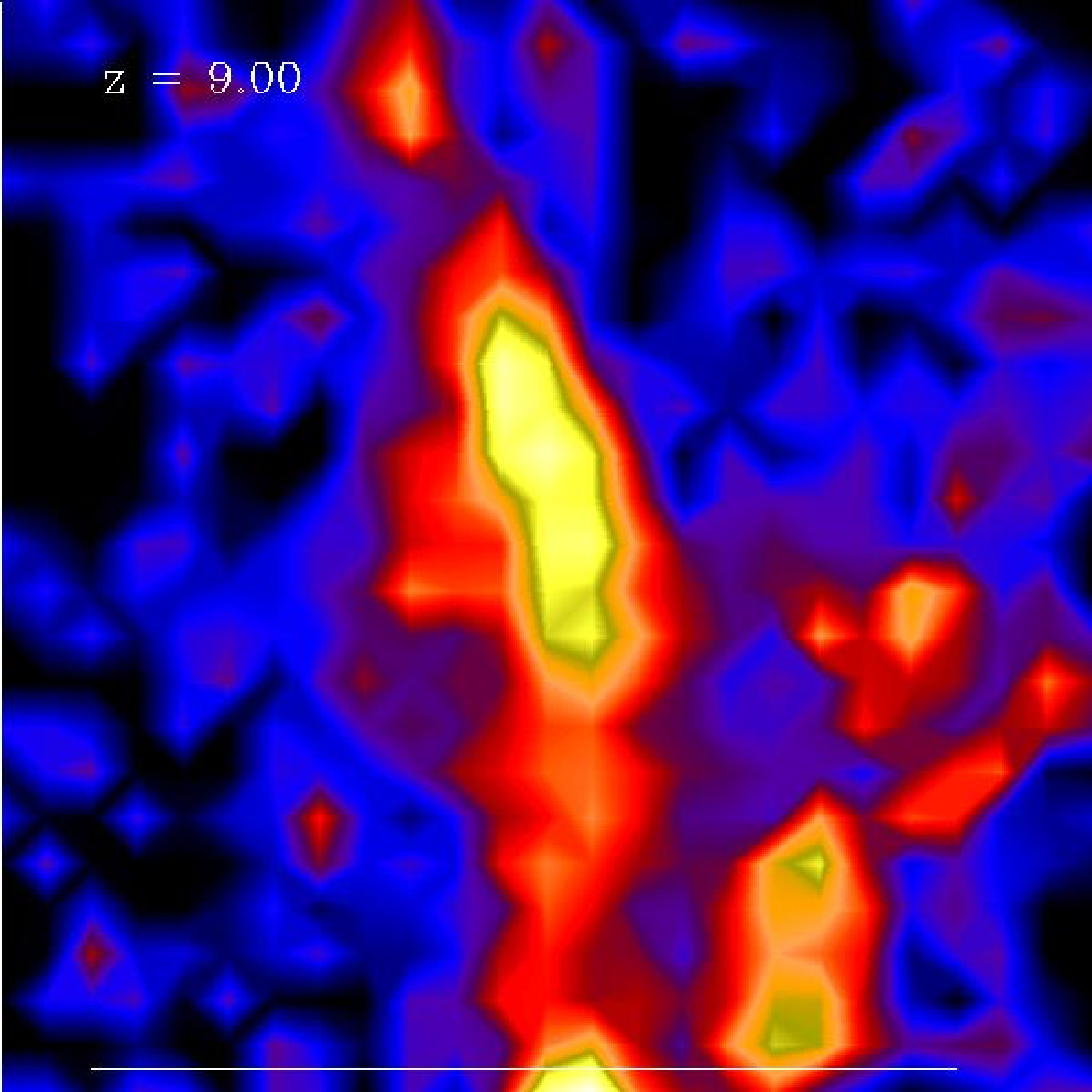}}
\end{center}
\caption{
\footnotesize
Snapshots of gas density (color-coded) evolution in the BDM model at $z=9$
(based on Romano-Diaz et al. 2008a). The frames are from left-to-right
600~kpc, 200~kpc, 50~kpc and 25~kpc across.
}
\label{fig1}
\end{figure*}

Here we shall take an alternative approach and treat an assembling disk--halo 
as an open system whose evolution is driven by both extrinsic (i.e., 
environmental) and intrinsic factors. In this context, `open' is synonymous 
with `cosmological.' The cosmological caveats include the 
redshift dependence of gas and dark matter (DM) influx along the large-scale 
filaments which feed the growing disk inside an assembling halo. Thus,
conditions which are associated with a disk-halo growth are those
of an asymmetric background DM potential and clumpy mix of DM and baryons.
This asymmetry of the potential comes from the surrounding
filaments(s) and from virialized triaxial DM mass distribution.
The accretion process itself is highly inhomogeneous and potentially 
responsible for various dynamical phenomena in the disk and its immediate 
surroundings --- few of these will be discussed here.

Galactic bars appear as dominant morphological features in nearby disks (e.g., 
Sellwood \& Wilkinson 1993; Knapen et al. 2000; Marinova \& Jogee 2007). At 
intermediate redshifts of $z\sim 0.2-1$, this picture 
is a subject to controversy (Jogee et al. 2004; Elmegreen et al. 2004; but see Sheth 
et al. 2008). At even higher $z$ we know very little about the disk evolution. 
However, the importance of bars goes well beyond morphology ---
they serve as a fundamental channel of angular momentum redistribution in a 
galaxy. Hence bar formation and evolution should be tied to the overall
process of galaxy formation and evolution in the CDM universe. Here we 
attempt to pursue this argument.
 
\section{First Generation of Bars in Open Systems}

Cold stream-fed growth emerges as the dominant mode of DM and baryonic
assembly in massive galaxies (e.g., Dekel et al. 2008). In the following 
discussion, 
we rely on representative models which run from identical initial 
conditions at $z=120$ in $\Lambda$CDM Universe --- pure DM (PDM) 
simulations and simulations with WMAP3 fraction of baryons, the BDM models 
(Romano-Diaz et al. 2008a). Fig.~1 displays snapshots of gas density 
evolution in the BDM model on scales from 600~kpc to 25~kpc. The latter
scale is relevant for disk formation and it is significant that the 
filamentary structure extends down to this size at high $z$.  The
resulting halo shapes will be substantially triaxial (e.g., Allgood
et al. 2006), partly because of the assembly process and partly as 
a result of the radial orbit instability. Baryons act as to wash out the 
equatorial ellipticity and to reduce the flatness of the DM distribution 
(Fig.~2; also e.g., Kazantzidis et al. 2004; Berentzen \& Shlosman 2006; 
Shlosman 2007).

\begin{figure*}[t!]
\begin{center}
\resizebox{13.8cm}{!}{\includegraphics[clip=true]{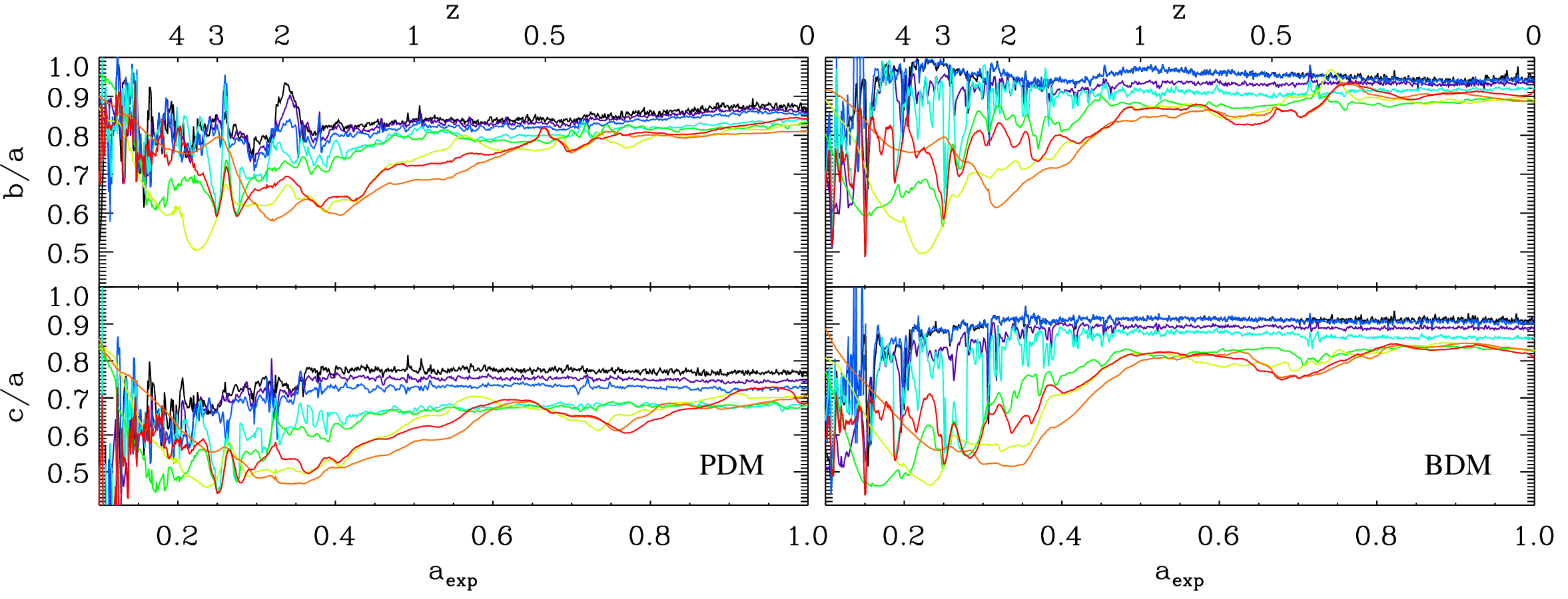}}
\end{center}
\caption{\footnotesize
Evolution of the axial ratios (shapes) for the prime DM halo at various
representative radii in the PDM (left) and BDM (right) simulations. Top:
halo $b/a$ (stratified roughly from top to bottom) at $r=10$~kpc, 20~kpc,
the NFW radius, radius of maximal circular
velocity, 100~kpc, 200~kpc, 300~kpc and the virial radius. Bottom: the
same for $c/a$ (Romano-Diaz et al. 2008a).
}
\label{fig2}
\end{figure*}

One of the
crucial differences between the local and high-$z$ universes lies in that the
latter is much more cold gas-rich. Gas-dominated disks that form in the 
asymmetric potentials cannot maintain axial symmetry --- stars and 
especially gas
will respond dramatically producing a standing density wave and a shock, 
respectively. Simulations have demonstrated that early bars will 
form under these conditions and that these bars will have properties
different from the canonical bars discussed in the literature so far 
(e.g., Heller et al. 2007b). Specifically, this first generation of
bars (1) will form in disks with a low surface density and central mass 
concentration. Consequently, (2) they will exhibit low pattern speeds. 
(3) They will experience a rapid growth, concurrently with that of the
host galactic disks. Finally, (4) both bars and disks will be aware of
the DM halo orientation, i.e., position of its major axis. Importantly,
the halo figure tumbling is exceedingly slow, whether it grows in a 
filament or in the field (Bailin \& Steinmetz 2004; Heller et al. 2007b;
Romano-Diaz et al. 2008a,c). This will result necessarily in a double
{\it inner} Lindblad resonance (ILR) --- the outer ILR will move far
out and the inner ILR to the center --- between these resonances the baryonic
response in the disk will be orthogonal to the major axis of the halo,
reducing its equatorial ellipticity.
 
Thus, first bars are expected to form in the growing disks embedded in the
triaxial halos. However, bars and triaxial halos are incompatible ---
a canonical bar tumbling within a stagnating halo will serve as a source
of chaos and dissolve quickly (El-Zant \& Shlosman 2002). The caveat lies
in that the early bars are supposed to rotate very slowly (see above) 
and so the amount
of chaos will be greatly reduced, prolonging the bar dissolution time
to $\sim 1-2$~Gyr. During this time period, the disk growth can be 
appreciable. If the disk response to the halo forcing (see above)
can successfully wash out the inner halo equatorial ellipticity over this
timescale, it will remove the main source of chaos and open the
possibility for the bar to strengthen. Future work can confirm
whether this indeed what is happening.

\begin{figure*}[]
\begin{center}
\resizebox{11.5cm}{!}{\includegraphics[clip=true,angle=-90]{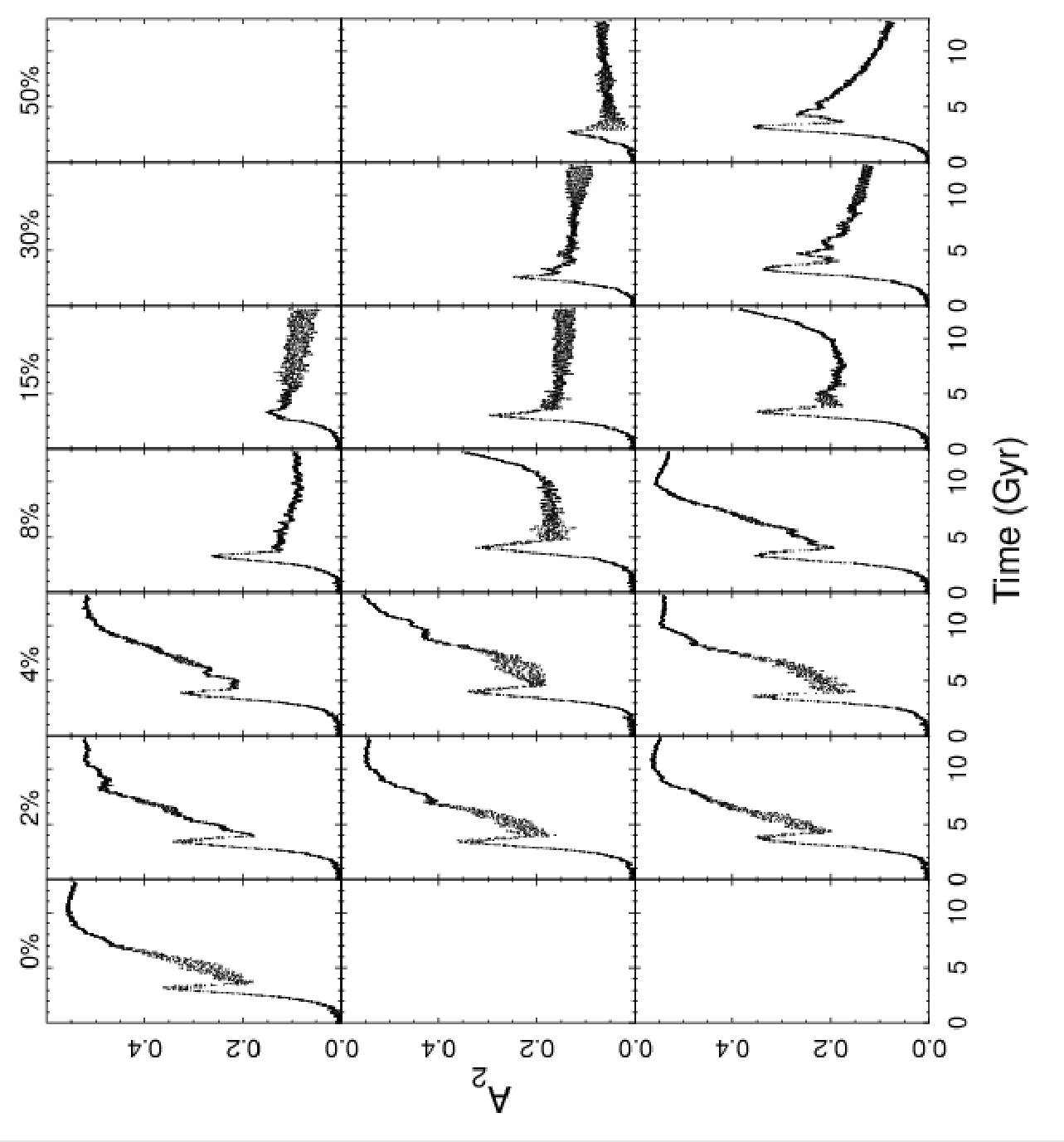}}
\end{center}
\caption{
\footnotesize
Evolution of the stellar bar strength, given by the Fourier amplitude
$A_2$ of the $m=2$ mode, for various gas fractions, $f_{\rm g}\sim 0\%-50\%$
in the disk immersed in the live halo. The time is given in Gyrs. The Jeans
instability is progressively damped from top to bottom models. The first
peak defines the time of the vertical buckling in the bar. Note that $A_2$
shapes are very similar in all models up to the first peak (Villa-Vargas et
al. 2008).
}
\label{fig3}
\end{figure*}

Bars living in growing disks are necessarily gas-rich (e.g., Heller et al.
2007a,b; Romano-Diaz et al. 2008a,c), to the extent 
that was not contemplated in the literature so far. Shlosman \& Noguchi 
(1993) found that, for gas fractions $f_{\rm g}\gtorder 10\%$ of the disk 
mass, the Jeans instability will lead to massive clumps which will spiral 
down to the central kpc as a result of a dynamical friction, leading to
a nuclear starburst. Heating the stellar disk on their way,
the clumps can damp the canonical bar instability or postpone it 
substantially. Bournaud
et al. (2005) claimed that $f_{\rm g}\sim 7.25\%$\  is sufficient to trigger
the bar dissolution due to the $J$ transfer from gas to stars in the disk,
in  simulations with a rigid halo. 
Berentzen et al. (2007) compared the evolution of disks with 
$f_{\rm g}\sim 0\%-8\%$, concluding that the apparent decrease in the bar
strength is not related to the gas content but rather to the vertical
buckling instability (e.g., Combes et al. 1990; Raha et al. 1991) in the bar, 
which weakens it substantially but allows for a secular rebuilding 
(e.g., Martinez-Valpuesta et al. 2004; 2006). Bars that are capable of
growing typically appear as `fast' bars, i.e., extending to near corotation, 
even in cuspy halos (Dubinski et al. 2008).

Gas fractions to be encountered at high $z$ can be an order of magnitude higher
than quoted above.
Results, based on the equilibrium models, show that  
dependence of the bar strength on $f_{\rm g}$ is not so dramatic. As long as
the Jeans instability in the gas is damped, $J$ transfer from the gas to the stars
is compensated by $J$ lost by the stellar bar to the DM halo. Fig.~3
demonstrates that development of the canonical bar instability is nearly 
independent of the gas fraction, 
up to $f_{\rm g} \sim 50\%$ at least, {\it if the Jeans instability in the gas 
is suppressed} 
(Villa-Vargas et al. 2008). The buckling proceeds similarly independent of 
$f_{\rm g}$. But the 
subsequent evolution of the bar strength differs --- while 15\% gas still 
allows for the secular growth of the bar, 30\% reduces the bar strength by a 
factor of 2--3 over the Hubble time.

\section{Dynamics of Bars in Open Systems} 

When gas rich bars are triggered by the asymmetric DM potential, they can appear
simultaneously on more than one spatial scale. By this we mean that bars on 
different scales will either have different pattern speeds or will be
situated at different position angles (in fact, orthogonal to each other) and
have the same speed. Most frequently they form
on smaller scales of a few$\times 100$~pc --- so-called nuclear bars, and
on the scales of a few kpc, so-called prime bars, although in some cases only 
one bar forms 
(Heller et al et al. 2007a,b). As we have stated above, {\it these bars do not
form as a result of the canonical bar instability but rather constitute the
disk response to the finite external perturbation}. Unlike most of the bars
in the local universe, their young counterparts appear to be delineated by star 
formation which is present not only close to the ILR(s) but
throughout the area of the bar. In a way they resemble inclined starbursting
disks, unless distinguished kinematically. 

As a first step, we discuss the behavior of single bars formed under these 
conditions, then comment on the evolution of double bars.
Bar dynamics can be studied by inspecting their pattern speed, $\Omega_{\rm b}$,
evolution. Here again the `cosmological' bars differ from their low $z$ 
gas-poor
representatives. Those that are triggered by the asymmetric potential are 
characterized by strongly growing $\Omega_{\rm b}$, concurrently with the 
disk growth. An alternative way of triggering bars, also by a finite perturbation, 
was known for some time as a tidal triggering (e.g., Byrd et al. 1986; Noguchi 1987), 
but was never considered to be important enough to explain the observed bar fraction
in  the universe. As we discuss below and in the next section, within a somewhat 
modified framework, this picture may become more appealing.

It is important to emphasize here that the class of tidally triggered bars
exhibit initial $\Omega_{\rm b}$ which are determined by the orbital speeds and
the impact parameters of the perturbers. But the subsequent evolution of 
$\Omega_{\rm b}$ in all three types of bars (namely, canonical, triggered
by the DM asymmetry or triggered tidally) is determined by the balance of 
angular momentum in the bar and by the central mass concentration, i.e., 
the shape of the inner rotation curve in a galaxy. Here, the role of gas,
the only dissipative agent, cannot be overestimated. Either
dynamically or secularly, the gas influx to the center contributes to the mass
growth there and so, even under conservation of $J$ in the bar, will cause an 
increase in  $\Omega_{\rm b}$. The net effect on the bar tumbling can be
estimated from the amount of $J$ supplied by the disk gas to the bar within its 
corotation radius, the growth of the central mass concentration, and the
amount of $J$ lost by the bar to the outer disk and especially to the DM halo. 

Simulations of isolated disks with an ad hoc addition of gas (Bournaud \& Combes
2002) and cosmological simulations of growing disks embedded in large-scale
filaments (Romano-Diaz et al. 2008c) demonstrated that bars can accelerate over 
a prolonged time period of a few Gyrs, during which the surface density of the disk, 
especially of its gas component, grows. This underlines the importance of
the cold gas influx via filaments. With this, the strength, $A_2$, of 
the cosmological bars (anti)correlates with $\Omega_{\rm b}$
--- weakening bars accelerate, and vice versa, as shown first for the pure stellar
bars in equilibrium models (Athanassoula 2003). Exceptions include the time periods 
of interactions between the developed bars and flybys, and the double bar systems 
(Romano-Diaz et al. 2008c).

In the case of double bars, only the nuclear bars accelerate after formation,
while the primary bar pattern speeds are quite steady over long time periods
(Heller et al. 2007a). Both shapes and pattern speeds of nuclear bars vary 
with their orientation to the prime bar (Heller et al. 2001; Shlosman \& Heller
2002; Englmaier \& Shlosman 2004; Juntai \& Debattista 2007) and this is visible
in models of cosmological double bars as well (Heller et al. 2007a). The
latter work has also shown that double bars can lock their pattern speeds
at various ratios of nuclear-to-primary $\Omega_{\rm n}/\Omega_{\rm p}$ ---
an issue of mode coupling.
 
\section{Bar Triggering by DM Substructure}

\begin{figure*}[t!]
\begin{center}
\resizebox{6cm}{!}{\includegraphics[clip=true]{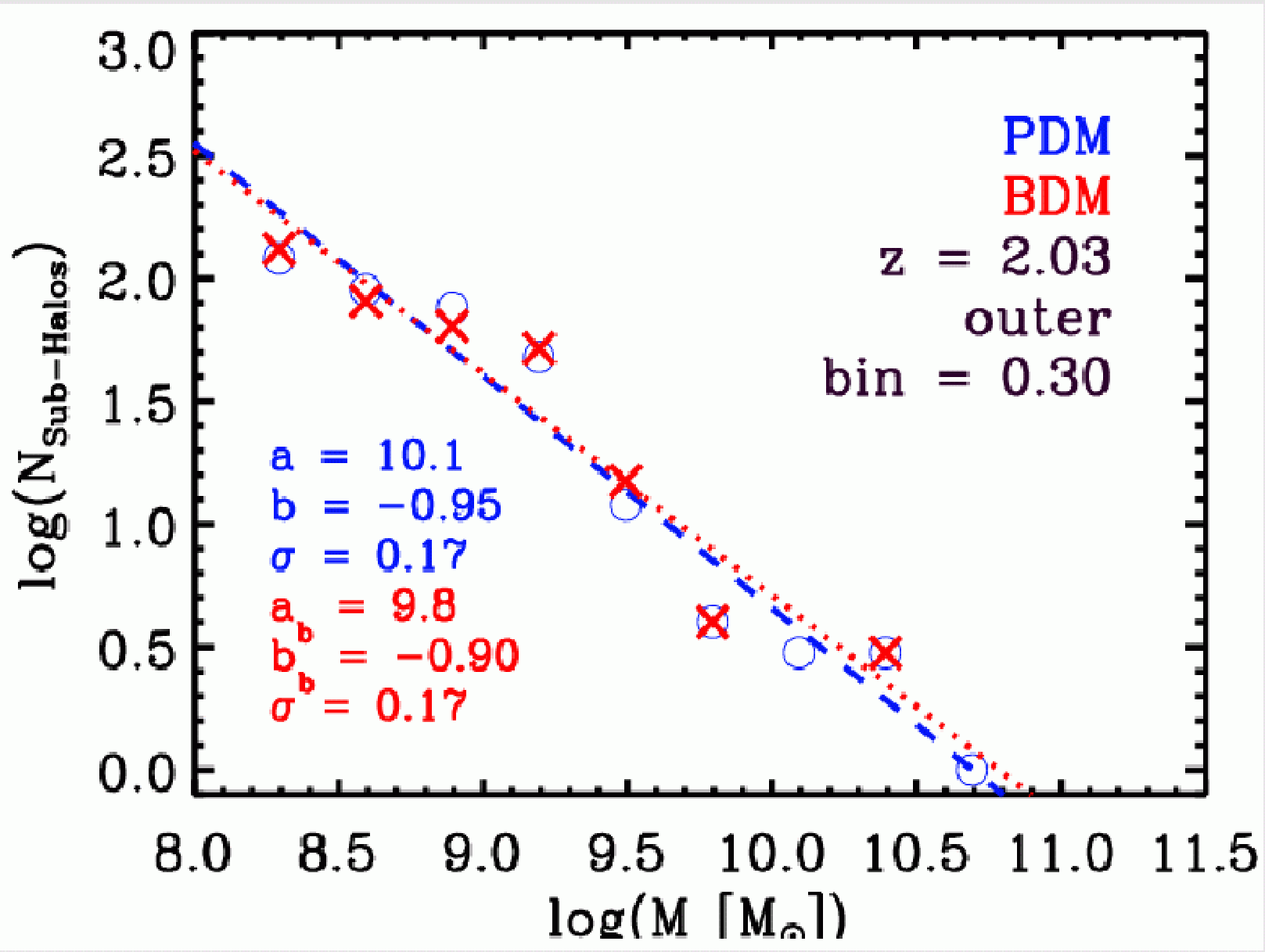}}
\resizebox{6cm}{!}{\includegraphics[clip=true]{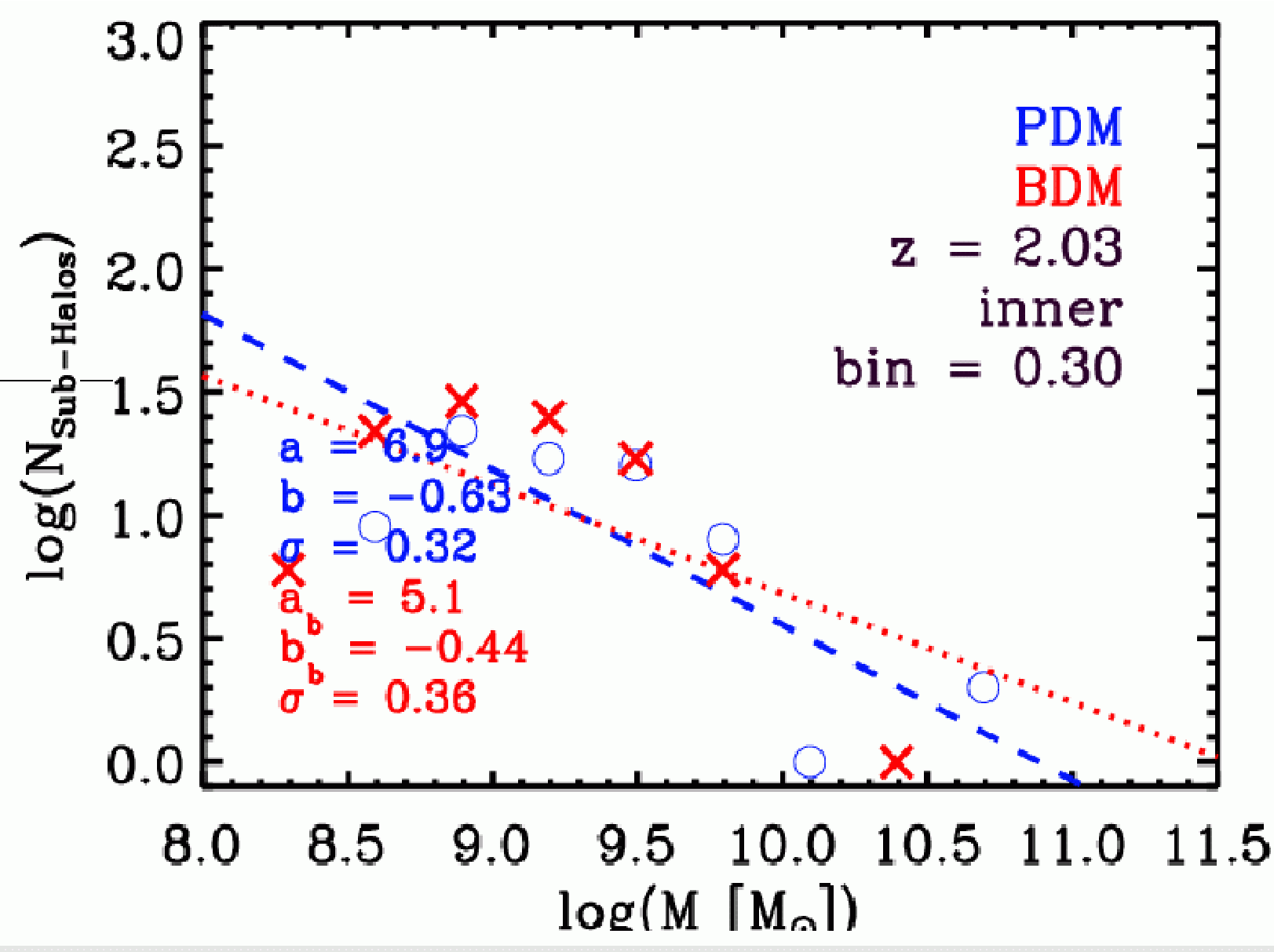}}
\end{center}
\caption{\footnotesize
Mass function $N{(M_{\rm sbh})}$ of the subhalo population at $z=2$ in PDM 
(blue dashed) and BDM (red dotted) models.
The minimal resolved subhalo mass is $10^8~{\rm M_\odot}$, which is about 
$10^{-4.3}$ of the prime halo virial mass at $z=0$. 
{\underline{\it Left frame:}}  
$N{(M_{\rm sbh})}$ for subhalos outside the virial radius of the prime halo;  
{\underline{\it Right frame:}} $N{(M_{\rm sbh})}$ for subhalos inside the 
central $\sim 100~{\rm kpc}\times 50~{\rm kpc}$ virialized region (defined by the 
isodensity contours) of the prime (Romano-Diaz et al. 2008d).
}
\label{fig4}
\end{figure*}

The possibility of bar triggering via interactions with other galaxies hints 
at a strong dependence
of bar fractions on the environment and redshift. However, bar fractions in 
clusters of galaxies and in the field appear comparable (Marinova et al. 2008; 
and these Proceedings). This discrepancy between expectations and observations 
can be erased if bars can be triggered by interactions with DM substructure and 
not by a more rare interactions with their more massive neighbors. Existence
of such DM subhalos is inherent to the hierarchical scenario of galaxy formation 
(e.g., White \& Rees 1978). Hence, whether a halo is found in the cluster or 
in the field, it will grow by consuming the surrounding subhalos --- those
that survive the radial plunge can have a profound effect on the disk evolution. 

\begin{figure*}[]
\begin{center}
\resizebox{6cm}{!}{\includegraphics[clip=true]{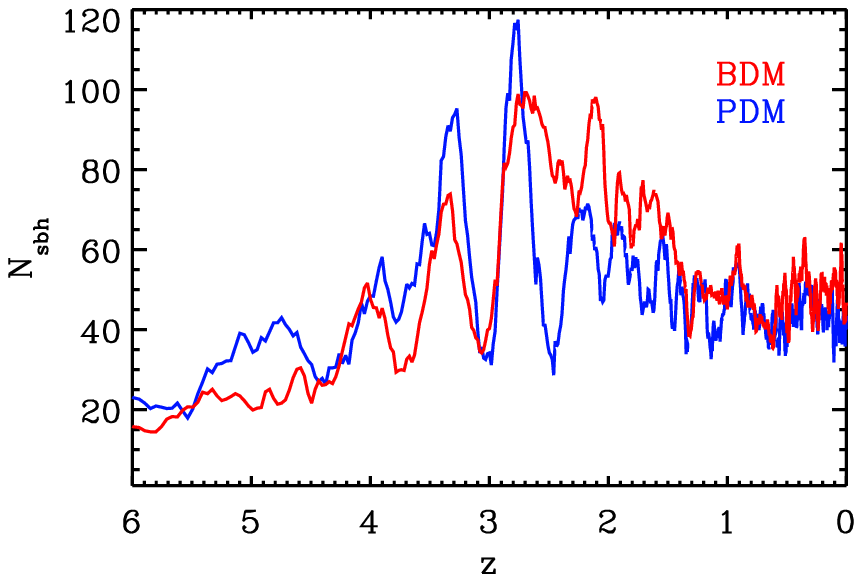}}
\resizebox{6cm}{!}{\includegraphics[clip=true]{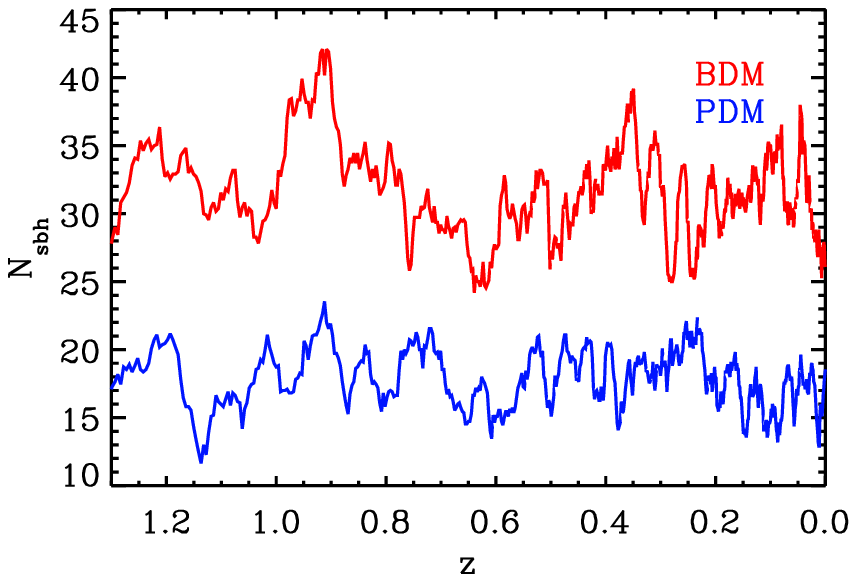}}
\end{center}
\caption{\footnotesize
Population of subhalos in PDM and BDM models within the inner 60~kpc
(left) and inner 30~kpc (right) of the prime halo at lower redshifts 
(Romano-Diaz et al. 2008b).
}
\label{fig5}
\end{figure*}

Properties of a subhalo population have been studied in pure DM simulations
(e.g., Ghigna et al. 1998; Klypin et al. 1999; Kravtsov et al. 2004; 
Diemand et al. 2007). Their mass distribution function was found to be
nicely approximated by a power law $N_{\rm sbh}\sim M_{\rm sbh}^{-1}$,
where $M_{\rm sbh}$ is the mass of a subhalo (e.g., Diemand et al. 2004).
Much less is known about subhalos in the presence of baryons. Their power
law distribution seems to be retained for DM halos above 
$10^{10}~{\rm M_\odot}$ --- the resolution limit (Weinberg et al. 2008).
Fig.~4 displays the subhalo mass functions for PDM and BDM models at 
$z\sim 2$, for much higher resolution simulations resolving subhalos above 
$10^8~{\rm M_\odot}$
(Romano-Diaz et a;. 2008d). To emphasize the environmental effects, the 
population was divided into the outer one, situated in the computational
box {\it outside} the virial radius of the prime halo, and the inner one, 
situated {\it within} the central virialized region delineated by density 
isocontours, 
$\sim $100~kpc~$\times $~50~kpc. A number of trends can be seen: (1) for the
`field' subhalos, the difference between the PDM and BDM models is minimal
at these redshifts.
(2) the log-log slope is about $(-0.9)-(-0.95)$ and quite stable at other
$z$. (3) Evolutionary trends are present in the inner subhalo population ---
their integrated number is lower by a factor of a few, both in the PDM and BDM.
(4) the BDM slope is shallower than the PDM one --- more massive BDM subhalos
survive better than the PDM ones in the central dense region of the prime halo.

\begin{figure*}[]
\begin{center}
\resizebox{3.3cm}{!}{\includegraphics[clip=true]{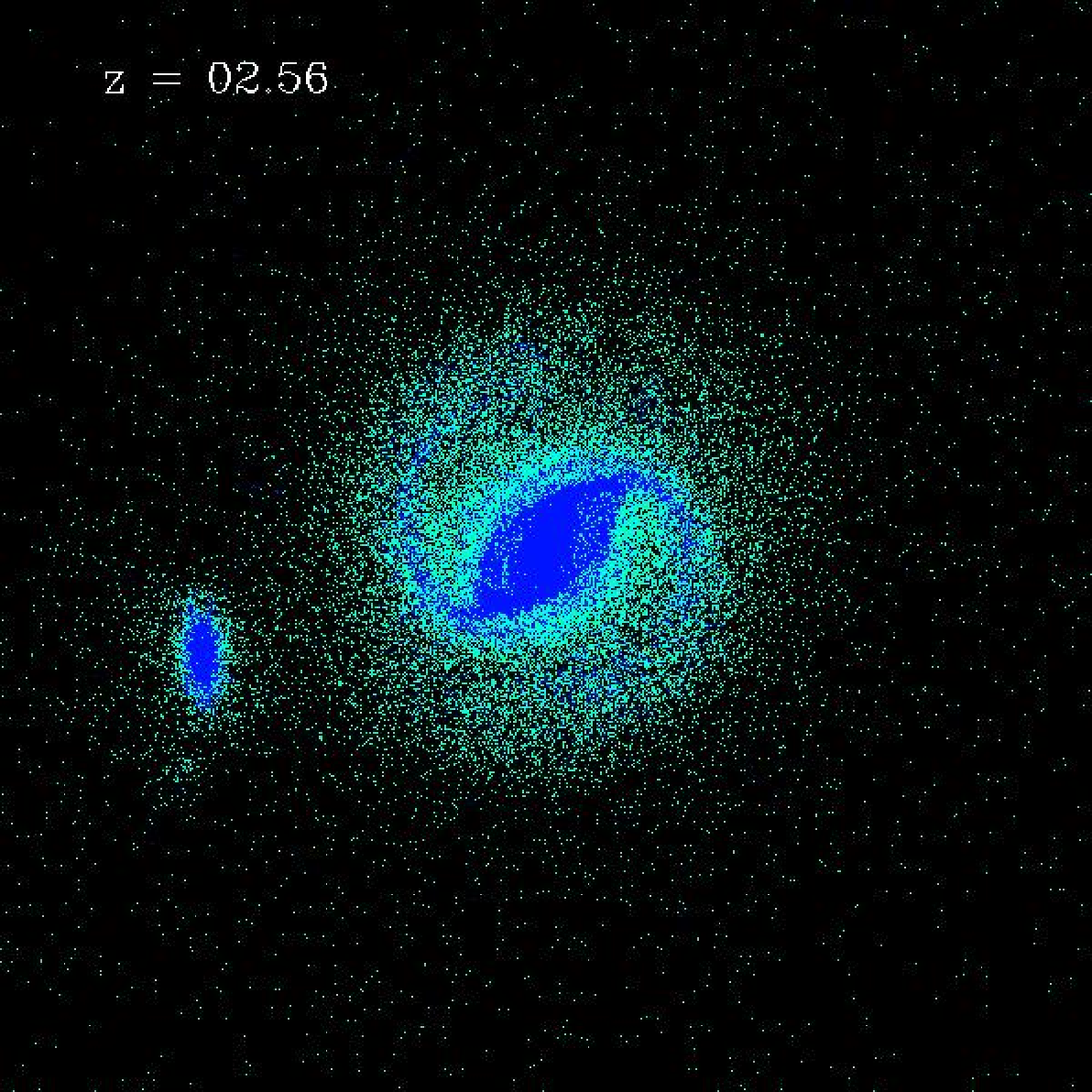}}
\resizebox{3.3cm}{!}{\includegraphics[clip=true]{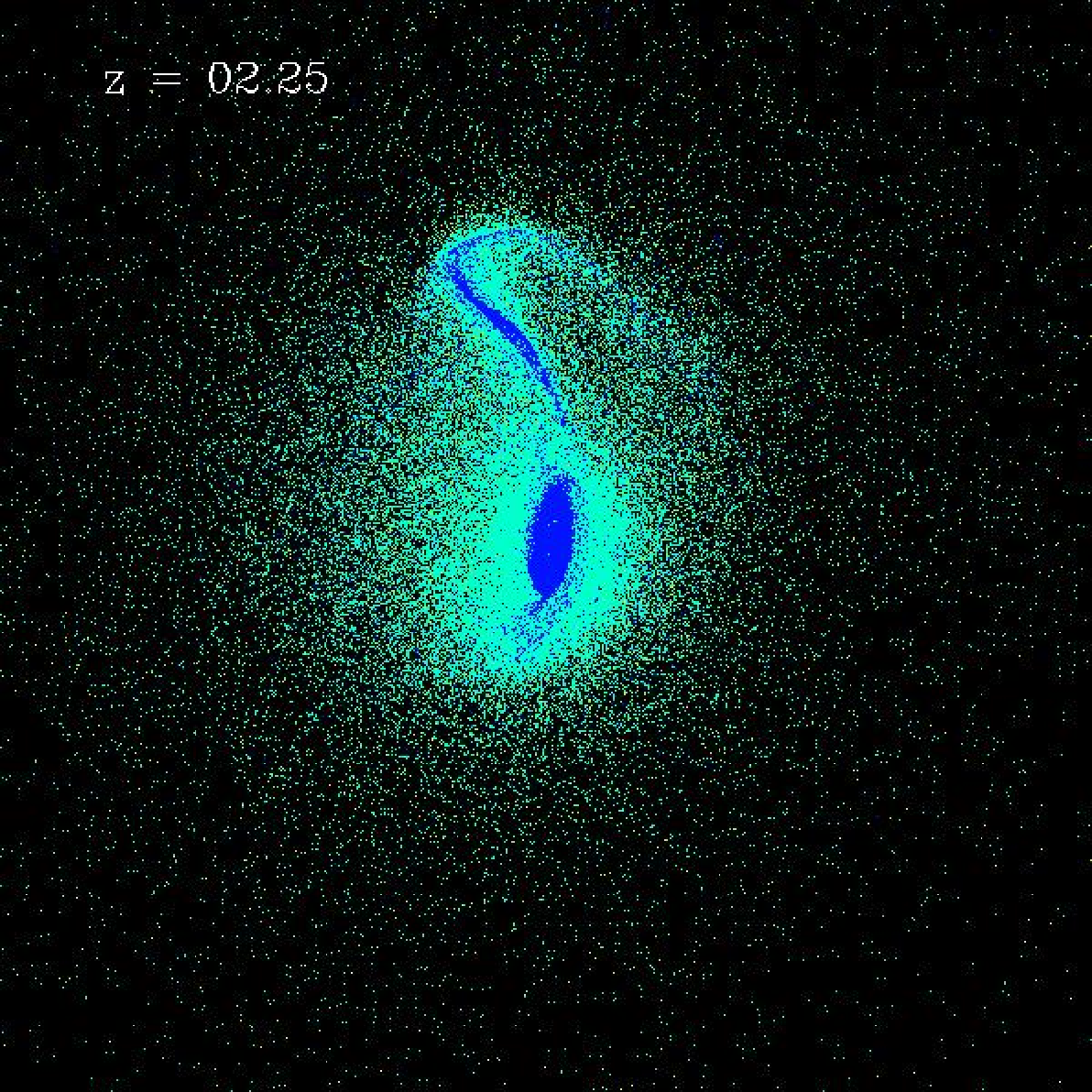}}
\resizebox{3.3cm}{!}{\includegraphics[clip=true]{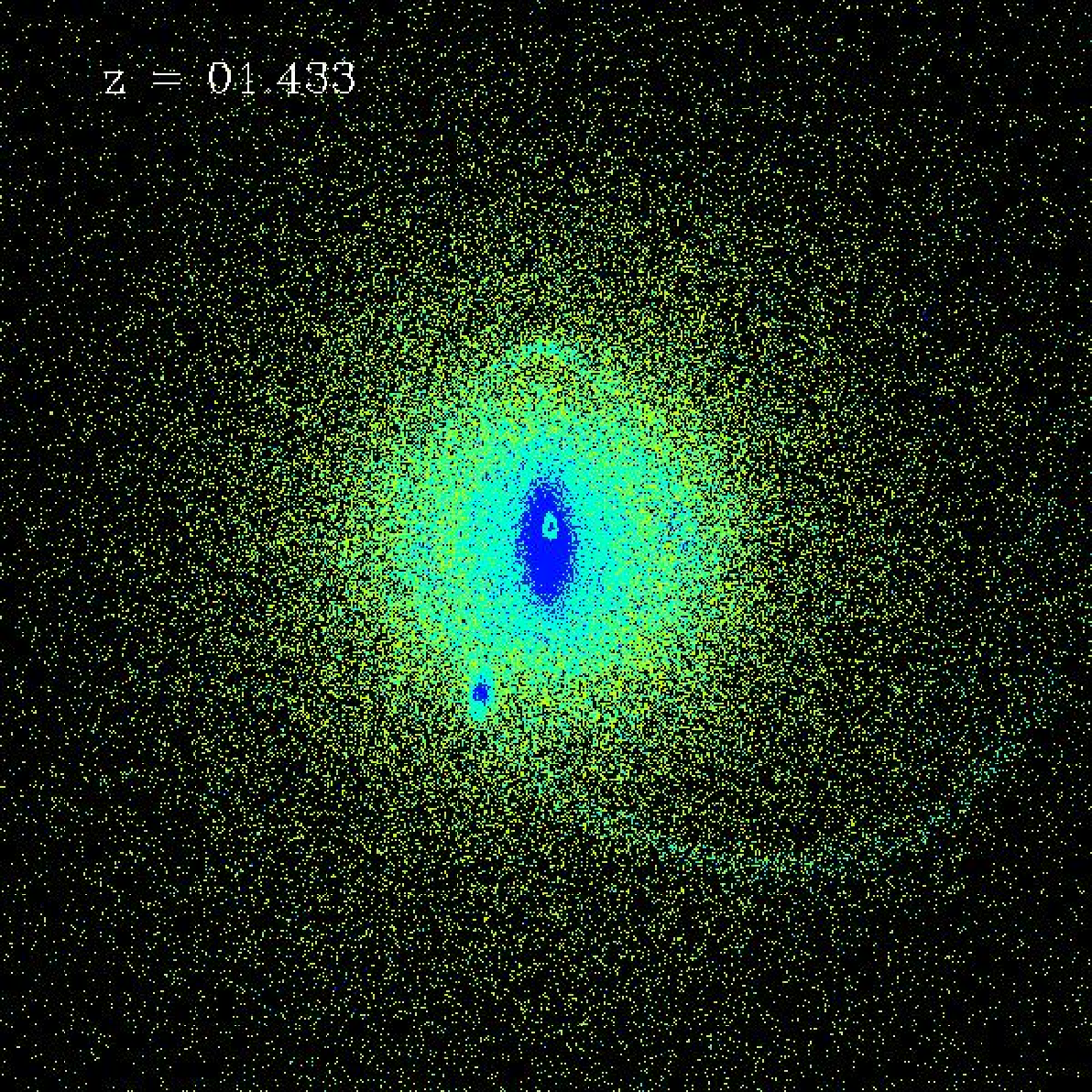}}
\resizebox{3.3cm}{!}{\includegraphics[clip=true]{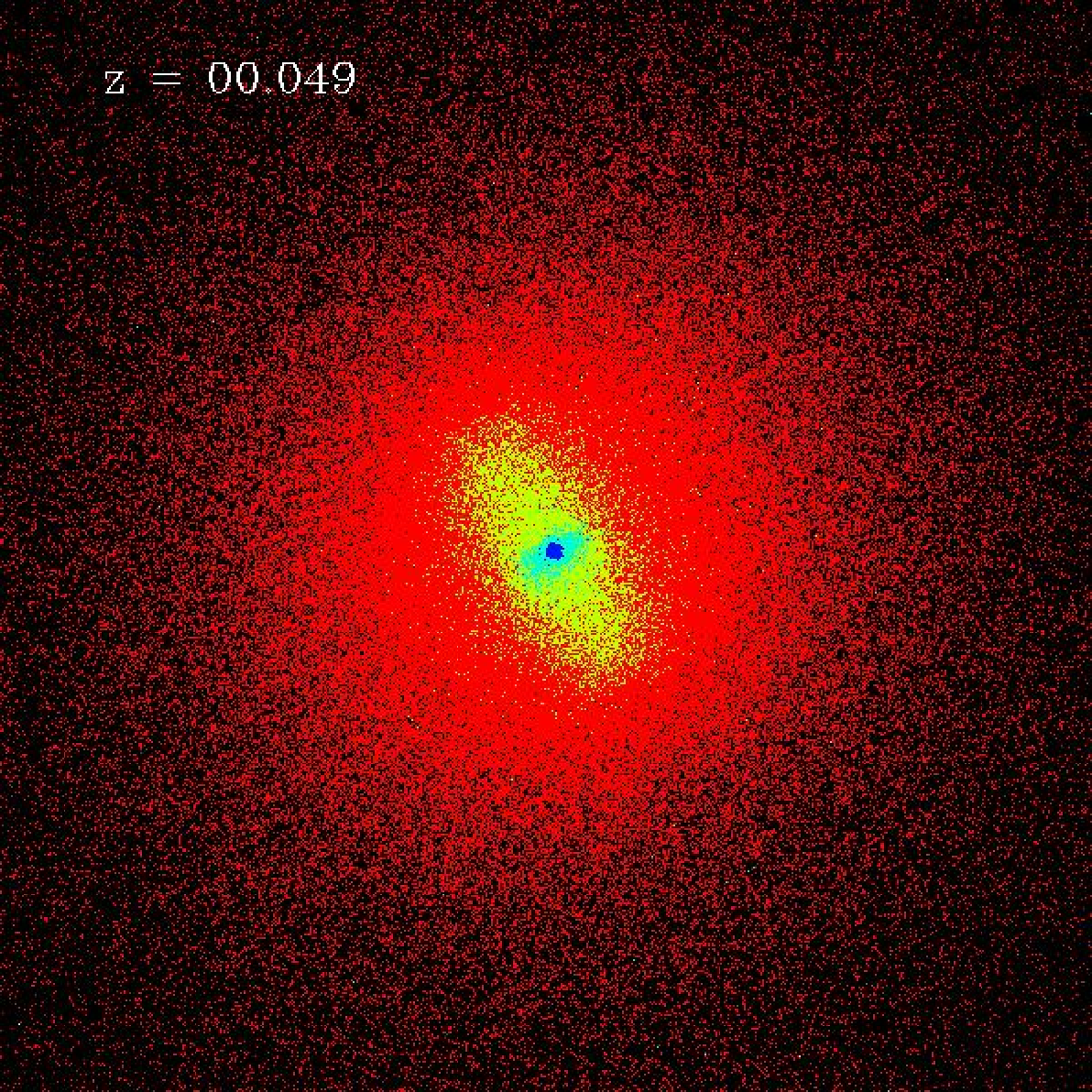}}
\end{center}
\caption{\footnotesize
Snapshots of disk evolution and its stellar population. Stellar ages are
0--0.2~Gyr (blue), 0.2--2~Gyr (blueish), 2--5~Gyr (yellow) and older than
5~Gyr (red). In this model, a long-lived stellar bar is triggered by
the interaction with an inclined prograde subhalo (hosting another disk)
at $z\sim 2.56$ (left). Numerous interactions with additional subhalos (e.g.,
middle frames) leave the bar intact. A nuclear bar developes after $z\sim 0.1$,
orthogonally to the prime bar (right). From Romano-Diaz et al. (2008c).
}
\label{fig6}
\end{figure*}
 
Our main interest in subhalo populations lies in its impact on the disk 
evolution. Specifically, we focus on the comparison between the PDM and BDM 
models. The environmental effects are clearly seen in the overall decrease
in $N_{\rm sbh}$, both with and without baryons. This trend accelerates
with $z$ and especially for the innermost massive PDM subhalos. Hence we expect
an excess of {\it inner} BDM subhalos over the PDM ones at lower redshifts.
Fig.~5 shows that indeed the BDM subhalos dominate that inner 30~kpc region
of the prime halo at low $z$, well after the epoch of major mergers which
ends by $z\sim 1.5$ in these models.

The time variable influx of the subhalos is clearly visible in Fig.~5. The 
maxima at $z\sim 4-2$
(left frame) are associated with the major mergers. The subhalo population
at $z\ltorder 1$ (right frame) is much more steady, still one can observe
`waves' of subhalos that penetrate the inner 30~kpc. A more careful analysis
shows that these subhalos have been accreted via filament(s). They
appear to cluster within the filament but their merging there
is suppressed. The ability of these subhalos to penetrate the central region
of the prime halo is enhanced because they are `glued' by the baryons and
hence can withstand larger tides. 

Interactions between the DM substructure and stellar bars can be classified
very roughly into flybys and direct hits (those include mergers). Prograde
encounters can excite bars, retrograde can only cause lopsidedness, while
direct hits weaken existing bars. Like the canonical bars formed as a result of the
intrinsic instability, tidally-induced bars are very resilient. They
avoid a number of difficulties associated with the linear stage of the bar
instability. An example
of such a long-lived bar triggered by a DM subhalo on a prograde inclined
orbit is shown in Fig.~6. Interactions with the surrounding substructure
leave it intact (Gauthier et al. 2006; Romano-Diaz et al. 2008c). Without 
much exaggeration,
one can say that in order to destroy this bar, one should largely destroy the
disk which hosts it, although it does go through phases when it
is quite weak.  

To summarize, there is a clear advantage to processes which involve tidal
triggering of galactic bars, at least at high to intermediate, and possibly
low $z$. This triggering can result either from an asymmetric DM distribution
or from interactions with the surrounding substructure. Such a triggering 
constitute a finite amplitude perturbation and is independent of a number problems
associated with the canonical (spontaneous) bar instability. Recent numerical
simulations involving baryons show a more dominant population of subhalos
in the innermost DM halo, in the regions which host growing galactic disks,
and after the epoch of major mergers. Finally, the availability of
substructure around field and cluster galaxies points to the bar fractions
being (reasonably) independent of the environment.

\begin{acknowledgements}
I am grateful to my collaborators on this issue,  
Emilio Romano-Diaz, Clayton Heller, Jorge Villa-Vargas, 
Lia Athanassoula, Ingo Berentzen, John Dubinski and Yehuda Hoffman. 
Partially funded by NASA, NSF and STScI grants.
\end{acknowledgements}

\bibliographystyle{aa}

\end{document}